ÉCOLE POLYTECHNIQUE
FÉDÉRALE DE LAUSANNE

# ON THE MINIMUM DISTANCE OF NON-BINARY LDPC CODES

RETHNAKARAN PULIKKOONATTU

ABSTRACT. Minimum distance is an important parameter of a linear error correcting code. For improved performance of binary Low Density Parity Check (LDPC) codes, we need to have the minimum distance grow fast with $n$, the codelength. However, the best we can hope for is a linear growth in $d_{\min}$ with $n$. For binary LDPC codes, the necessary and sufficient conditions on the LDPC ensemble parameters, to ensure linear growth of minimum distance is well established. In the case of non-binary LDPC codes, the structure of logarithmic weight codewords is different from that of binary codes. We have carried out a preliminary study on the logarithmic bound on the the minimum distance of non-binary LDPC code ensembles. In particular, we have investigated certain configurations which would lead to low weight codewords. A set of simulations are performed to identify some of these configurations. Finally, we have provided a bound on the logarithmic minimum distance of nonbinary codes, using a strategy similar to the girth bound for binary codes. This bound has the same asymptotic behaviour as that of binary codes.

Advisor : Iryna Andriyanova
Advisor : Rüdiger Urbanke

*Date*: 2008 June 19.

## Contents

## 1. INTRODUCTION

In this report, we present some findings of a preliminary investigation on the minimum distance growth of non binary LDPC codes. We present some aspects of the necessary conditions for the growth of minimum distance for non binary LDPC codes.

## 2. LOW-DENSITY PARITY-CHECK CODES

Low-Density Parity-Check codes (LDPC) are class of linear error correcting codes originally proposed by Gallager in the 1960s [1]. LDPC codes are among the capacity achieving codes, with superior performance under iterative decoding algorithms which are easy to implement with affordable complexity [2].

### 2.1. Binary LDPC Codes: Construction and parameters.

A linear binary code $\mathcal{C}$ of length $n$ is a linear subspace of $\mathbb{F}_2^n$. If $\mathcal{C}$ has dimension $k$, then $\mathcal{C}$ is referred as $\mathcal{C}[n,k]$ code [3]. The code $\mathcal{C}$ forms a linear map from all possible binary $k$ tuples to a $k$ dimensional vector space $\mathbb{F}_2^n$ over $\mathbb{F}_2$. A codeword $c \in \mathcal{C}$ is an element in the vector space $\mathbb{F}_2^n$. The linear mapping is represented by a generator matrix $G \in \mathbb{F}_2^{n\times k}$. Being a linear subspace of dimension $k$, the code $\mathcal{C}$ can also be described as the kernel of a matrix $H \in \mathbb{F}_2^{(n-k)\times n}$, so that $\mathcal{C} = \{c \in \mathbb{F}_2^n | Hc = 0\}$ (We treat codewords $c$ as column vectors for this description). The matrix $H \in \mathbb{F}_2^{(n-k)\times n}$ is called parity check matrix. The generator and parity check matrices are related by $HG = 0$.

LDPC codes belong to the class of linear codes. A LDPC code is defined as follows [4].

**Definition 1.** A low density parity check code is a linear block code which has a sparse parity check matrix.

Here, sparse refers to the condition of having at most a fixed constant number of 1's (the constant is independent of $n$) [5]. The name *low density* in LDPC is attributed to this sparsity feature of the parity check matrix.

The number of ones in a binary vector is referred as its weight. For a matrix, row (column) weight denote the weights of the corresponding row(column) vector. A LDPC code is regular, if all the column weights are the same (say $l$) and all the row weights (say $r$) are the same, for the parity check matrix $H \in \mathbb{F}_2^{m\times n}$ (Note that, the number of parity equations are usually, denoted as $m$ instead of $(n-k)$ as described for linear codes in general. We adopted this commonly used notation in LDPC literature [2]). This represent a regular $(n,l,r) - LDPC$ code. The relationship $ln = rm$ holds. The parameters $l(r)$ are also known as variable(check) node degrees for $(n,l,r) - LDPC$ code.

The *design rate* of a $(n,l,r) - LDPC$ is defined $R = 1 - l/r$. A toy example, of a regular parity matrix with $n = 10$, $l = 3$ and $r = 6$ is represented below.

$$\begin{bmatrix} 1&1&1&0&0&1&1&0&0&1\\ 1&0&1&0&1&1&0&1&1&0\\ 0&0&1&1&1&0&1&0&1&1\\ 0&1&0&1&1&1&0&1&0&1\\ 1&1&0&1&0&0&1&1&1&0 \end{bmatrix}$$

When the column (row) weights are not identical across columns (rows), such matrices correspond to what is known as left (right) irregular LDPC codes. We can describe them in terms of the column (row) weight distributions. An easier and more convenient representation using a graphical method is widely used, which we discuss next.

### 2.2. Tanner graphs.

Tanner introduced in [6] a convenient graphical representation of LDPC codes in terms of a bipartite graph. Such a representation also known as Tanner graph, is often useful in the encoding , decoding, as well as in the analysis of LDPC codes [2]. Bipartite representation and parity matrix representation are essentially synonymous. A bipartite representation of a simple parity check matrix Eq.(1) is illustrated in Figure.1.

$$\begin{bmatrix} 1&0&1&1&0&0&0&0\\ 1&0&0&1&0&1&1&0\\ 0&0&1&1&1&0&0&1\\ 1&0&0&0&0&1&1&1\\ 0&1&1&0&1&1&0&0\\ 1&0&0&0&1&0&1&1 \end{bmatrix} \tag{1}$$

A bipartite graph consists of a set of variable nodes and a set of check nodes, together with edges connecting pair of nodes, of different type (An edge neither connects a variable node to variable node, nor from a check node to check node). The variable nodes (shown as circular nodes in Figure.1) represent the elements (bits) of codeword $(x_1, x_2, \ldots, x_n)$ and check nodes (shown as rectangular nodes in Figure.1) represent $m$ parity equations. An edge $e_{i,j}$ is connected from a variable node $x_i$ to a check node $c_j$ if the element $H_{j,i}$ is 1 (nonzero). The number of edges connected to a variable (check) node is referred to as the *degree* of the corresponding variable (check) node. For regular $(n,l,r) - LDPC$, this result in $l(r)$ neighbours for all the variable (check) nodes. The column (row) indices of $H$ maps to the variable (check) nodes in bipartite graph. Thus, the weight of a column (row) simply equals to the degree of corresponding variable (check) node in bipartite graph.

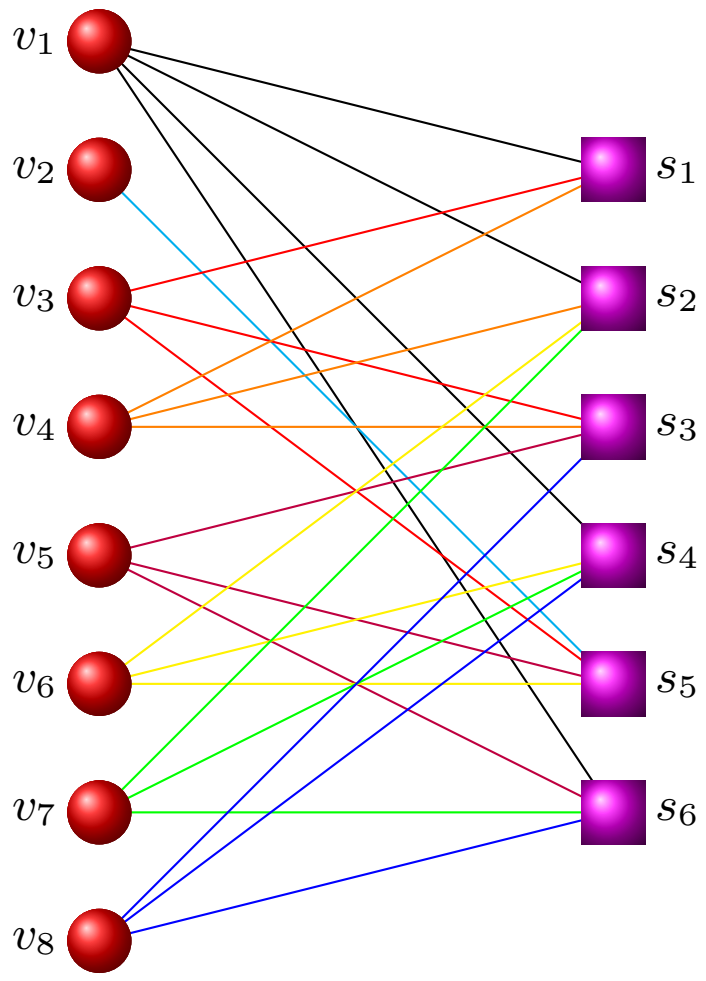

**Figure 1**. Bipartite representation a code: The circular nodes are called variable nodes and square boxes referred as check nodes representing parity conditions. The number of variable nodes equal the code length and the number of check nodes equal to the number of parity equations. In this illustration, the codelength $n = 8$ and the number of parity conditions $m = 6$. The graph maps to the parity check matrix $H$ in Eq.(1) where the row index correspond to the check nodes and the column indices to variable nodes. This is an example of irregular code.

An irregular LDPC code has a sparse parity check matrix in which the column (row) weight may vary from column (row) to column (row). In such cases, it is useful to talk about the distribution of the weights on column(row) of the $H$ matrix. The bipartite graph illustration in Figure.1 is an example of irregular code.

2.3. **Degree distribution pairs.** We have seen that, the weight of a column (row) of $H$ is the same as the degree of corresponding variable (check) node in the bipartite graph. For a general bipartite graph we could then define the degree distribution, which equivalently translate to the non zero elements of the representing parity check matrix. Let $L_i(R_j)$ denote the fraction of variable (check) nodes with exactly $i$ edges connected to them. Let $l_{\max}(r_{\max})$ denote the maximum number of edges connected to any variable (check) node. The polynomial $L(x) = \sum_i^{l_{\max}} L_i x^i$ is the variable node degree distribution and $R(x) = \sum_j^{r_{\max}} R_j x^j$ is the check node degree distribution. These are usually described as pair $(L, R)$ and is referred to as degree distribution pair from node perspective. There is an equivalent definition of degree distribution from edge perspective, denoted as $(\lambda, \rho)$. The polynomials are $(\lambda(x) = \sum_i \lambda_i x^{i-1}, \rho(x) = \sum_j \rho_j x^{j-1})$ where $\lambda_i$ is the fraction of edges which are connected to variable nodes of degree $i$. Similarly, $\rho_j$ denote the fraction of edges which are connected to check nodes of degree $j$. For a $(n, l, r) - LDPC$-regular code we have: $\lambda(x) = x^{l-1}$ and $\rho(x) = x^{r-1}$ or equivalently, $L(x) = x^l, R(x) = x^r$

2.4. **Ensemble of binary LDPC codes.** It is of interest to study the ensemble properties of LDPC codes, rather than that of an isolated code. For large $n$, the performance of any code is found to be close to that of an ensemble [2].

Given a pair $(\lambda, \rho)$ of degree distributions and the block length $n$, an ensemble of bipartite graphs $\mathbf{G}(\lambda, \rho)$ is defined by the collection of graphs by running over all possible permutations of edges connecting variable nodes and check nodes, subject to the given degree distribution pair $(\lambda, \rho)$.

In what follows, when we discuss properties and performance of LDPC codes, we always refers to that of LDPC ensemble.

## 3. Minimum distance of LDPC Codes

Minimum distance is an important design parameter for a linear code [3]. The definition of minimum distance for LDPC is the same as that of any linear code.

**Definition 2.** Minimum distance is the smallest Hamming distance between any two codewords of the code.

More precisely, it is the smallest (among all codeword pairs) number of difference in bit values at individual positions of any two codewords. Minimum distance is denoted by $d_{\min}$.

3.1. **Necessary condition to have linear growth in minimum distance.** The behaviour of the minimum distance of a LDPC code ensemble as the codelength $n$ increases is referred as *minimum distance growth*. For improved decoding performance, we wish to have $d_{\min}$ grow fast with $n$, the codelength [7] [8]. However, the best we can achieve is a linear growth of $d_{\min}$ with $n$. We want to construct LDPC ensemble with linear *minimum distance growth*.

In order to achieve this growth, we have to admit necessary and sufficient conditions on ensemble parameters, more precisely conditions on degree distribution pairs $(\lambda, \rho)$. If $(\lambda, \rho)$ satisfies the necessary conditions, we could achieve *minimum distance growth* better than sub-linear (more precisely speaking, logarithmic). However, necessary condition does not guarantee linear *minimum distance growth*. To ensure *minimum distance growth* faster than logarithmic, we must avoid codewords of logarithmic minimum distance.

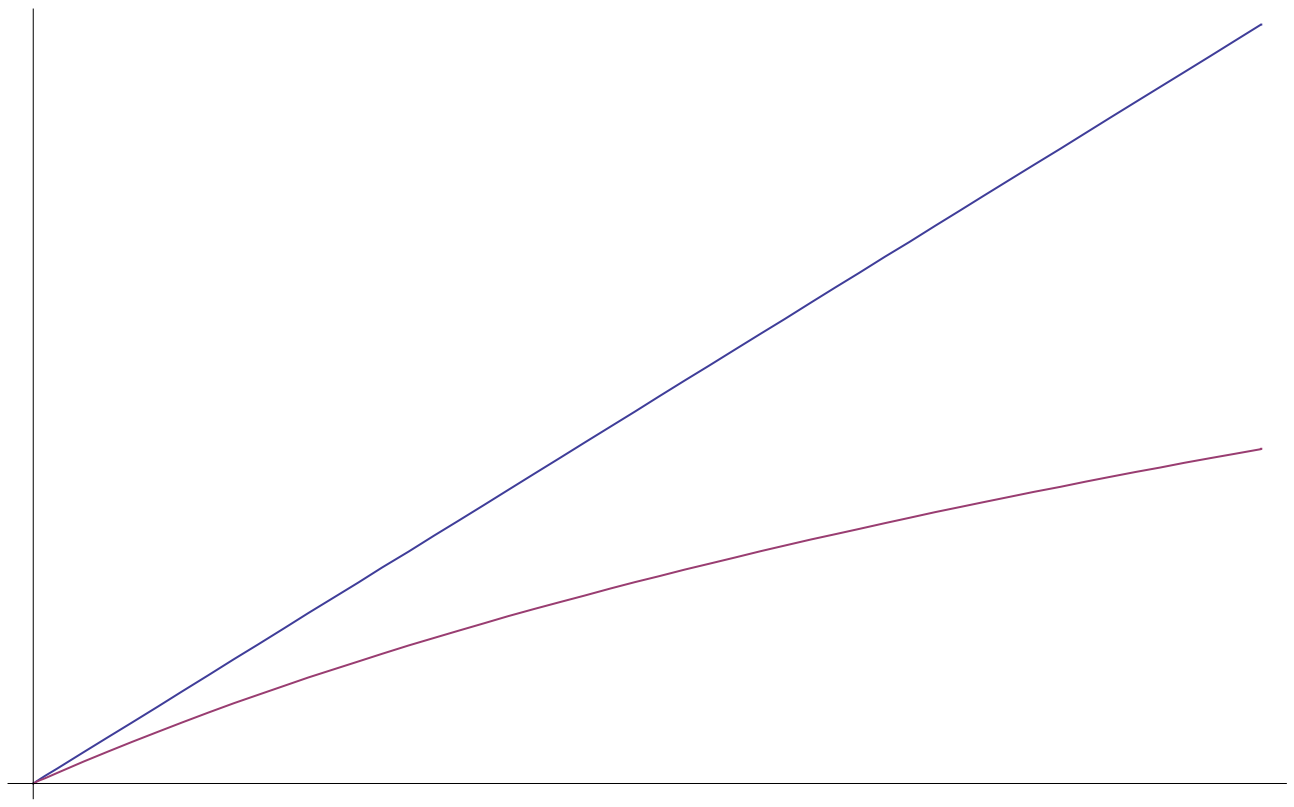

**Figure 2**. *Minimum distance growth* growth: The yaxis correspond to the growth in minimum distance, whereas, the xaxis show the codelength. The region between the two curves correspond to minimum distance growth faster than logarithmic and slower than linear. The bottom curve shows a logarithmic growth in minimum distance whereas, the top curve correspond to the linear minimum distance growth.

For binary codes, the necessary condition translates to the following:

If the number of degree 2 variable nodes ($n_2$) is equal to the number of check nodes $m$, each cycle correspond to a valid codeword.. When $n_2 \leq m$, the *minimum distance growth* is at most logarithmic in $n$.

## 4. Non-binary LDPC Codes

A non-binary LDPC code can be defined analogous to the binary LDPC code. These are linear codes defined as vector space $\mathbb{F}_q^n$ over $\mathbb{F}_q$, where $q$ is a prime power, i.e., $(q = p^a)$ with $p$ prime and $a \in \mathbb{N}$ a positive number. We will restrict attention only to cases where $q$ is of the form $2^a$,i.e., that is codes defined over extension fields of binary fields. A non-binary code denoted $\mathcal{C}[n,k]_q$ can be described as the kernel of the parity check matrix $H \in \mathbb{F}_q^{(n-k)\times n}$, such that $\mathcal{C} = \{c \in \mathbb{F}_q^n | Hc = 0\}$.

**Definition 3.** A low density parity check code is a linear block code, which has a sparse parity check matrix. The sparse elements are elements from $\mathbb{F}_q$. The code is defined over a finite field $\mathbb{F}_q^n$.

Similar to the binary code, a non binary code can be described using the Tanner graph representation of the corresponding parity check matrix. A non binary LDPC code defined over $\mathbb{F}_q$ with parity check matrix $H \in \mathbb{F}_q^{(n-k)\times n}$ has a $q-$ry bipartite graph representation. The edge labels, variable nodes ($q-$ry codeword $\in \mathbb{F}_q^n$) in $q-$ry bipartite graph assume nonzero values from $\mathbb{F}_q$. The linear operations (parity check equations) are as well carried out in the usual finite field algebra in $\mathbb{F}_q$. We provide an example of a $4-$ry bipartite graph representation (See Figure.3) of a parity check matrix defined over $\mathbb{F}_4$ shown below (Eq.2).

$$\begin{bmatrix} 1 & 0 & 1+\alpha & \alpha & 0 & 0 & 0 & 0 \\ 1+\alpha & 0 & 0 & 1+\alpha & 0 & \alpha & 1 & 0 \\ 0 & 0 & 1 & 1 & 1+\alpha & 0 & 0 & 1 \\ \alpha & 0 & 0 & 0 & 0 & 1 & 1+\alpha & 1 \\ 0 & \alpha & 1 & 0 & 1 & 1+\alpha & 0 & 0 \\ \alpha & 0 & 0 & 0 & 1 & 0 & \alpha & 1+\alpha \end{bmatrix} \tag{2}$$

In the example [2], the primitive polynomial generating $\mathbb{F}_4$ is $p(x) = 1 + x + x^2$. The primitive root is $\alpha$ and the elements of $\mathbb{F}_4$ are $\{0, 1, \alpha, 1+\alpha\}$.

4.1. **Code Ensemble.** The code ensemble of non-binary LDPC codes is defined in analogous way [2] as linear LDPC codes, with the exception that, the non-zero entries of the parity check matrix $H$ are chosen uniformly at random from $\mathbb{F}_q^\star$, where $\mathbb{F}_q^\star = \mathbb{F}_q \setminus \{0\}$. In words, the edge labels are chosen uniformly from the non-zero elements of $\mathbb{F}_q$, subject to the degree distribution pair $(\lambda, \rho)$.

4.2. **Minimum distance of non-binary LDPC codes.** For codes defined over $\mathbb{F}_q$ with $q = 2^a$, there are two possible definitions of minimum distance for non-binary LDPC codes (linear codes in general). One of them is the Hamming minimum distance, denoted by $d_{H\min}$. The Hamming minimum distance $d_{H\min}$ is simply the minimum (among among all codewords) number of difference in symbol positions between any two codewords in $\mathcal{C}$. The second definition of minimum distance is in terms of the binary image representation of the $q-$ry codeword. We adopt this latter definition of minimum distance for our investigation.

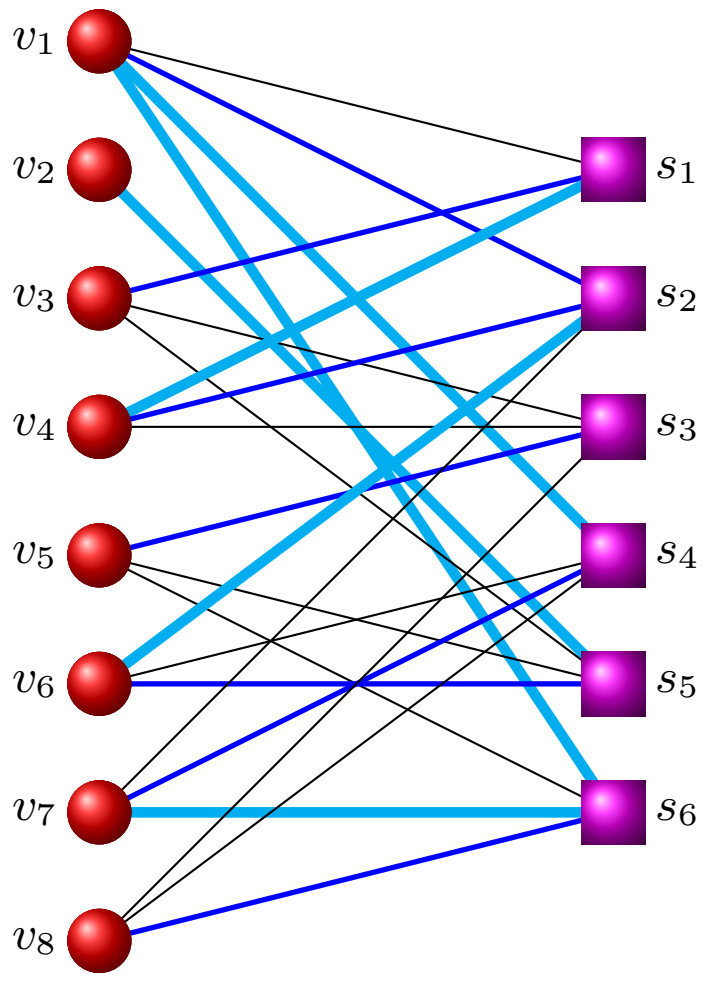

**Figure 3**. Bipartite graph of a non binary code with parity check matrix $H$ code defined over $\mathbb{F}_4$. The primitive polynomial generating $\mathbb{F}_4$ is $p(x) = 1 + x + x^2$. The primitive root is $\alpha$ and the elements of $\mathbb{F}_4$ are $\{0, 1, \alpha, 1 + \alpha\}$. The different edge labels are marked in different color (and thickness). Variable nodes are represented by circles and check nodes by square boxes.

**Definition 4.** Minimum distance of non-binary $q-$LDPC code is the smallest Hamming distance between the binary image representation of any two codewords.

## 5. Configurations for codewords with logarithmic weight

We look at the necessary conditions for the linear minimum distance growth of non-binary LDPC code ensembles. Our approach towards this direction is to study configurations which correspond to low weight codewords. More precisely, we focus on codewords which have logarithmic weight.

Consider a Tanner graph of binary LDPC code. We consider the sub graph induced by degree-2 variable nodes. Any cycle in the sub graph would correspond to a valid codeword. If length of the cycle is logarithmic, then it leads to logarithmic weight. It can be shown that when the number of degree 2 variable nodes in the code ($n_2$ is higher than the number of check nodes $(m)$, there exists cycles of logarithmic weight ($\lambda^{'}(0)\rho^{'}(1) < 2$)). So the necessary condition for better than logarithmic minimum distance growth is to ensure that, $n_2 \leq m$.

The structure of logarithmic weight codewords in non-binary case is however different from that of binary case. We have considered specific configurations involving cycles, union of cycles and chain of cycles. In the following section, we look at the structure of the sub-graphs induced by these configurations.

## 6. Consistency of structured equations in $\mathbb{F}_q$

Here, we focus on the structure of the sub-matrices (in the parity check matrix) corresponding to the sub-graphs of the configurations of interest. The system of linear equations involving these structured matrices will then provide some clues on the behaviour of codes satisfying the parity conditions.

First let us consider a linear system of equations in $\mathbb{F}_q$ (A general linear system of equations in finite field is presented in [10] and special structures of equations in $\mathbb{F}_2$ is addressed in [11]). We are specially interested in a system of the form $Ax = 0$, where, $A \in \mathbb{F}_q^{T_c \times T_v}$ and $x \in \mathbb{F}_q^{T_v}$. We associate the matrix $A$ to a bipartite graph $G$ with $T_v$ variable nodes and $T_c$ check nodes.

For a linear system of equations $Ax = 0$, the relationship between the rank criteria and the number of possible solutions in $\mathbb{F}_q$ is summarized in the following lemma.

**Lemma 5.** *If there are $T_v$ unknowns and $T_c$ equations ($T_v \geq T_c$. i.e., we have equal or more unknowns than equations), the number of solutions of $x \in F_q$ satisfying the linear system of equations $Ax = 0$ where $A \in \mathbb{F}_q^{T_c \times T_v}$ is equal to $q^{T_v - rank(A)}$.*

If the matrix $A$ is square ($T_c = T_v$), then, there are $q^{T_c - rank(A)}$ solutions. When $A$ is full rank ($rank(A) = T_c$), we have a unique solution and this unique solution is the all zero vector $0^{T_v \times 1}$.

Suppose we focus our interest to system of equations with special structure. More specifically, we look at systems with more unknowns than equations ($T_v > T_c$). Let $t = T_v - T_c + 1$. We consider matrix $A$ to have full rank $rank(A) = T_c$. We also consider the system (of equations and solutions) constrained such that, each elements of them are non zero. Stated differently, this means that the elements of the $x$ are all from $\mathbb{F}_q^{\star}$, where $\mathbb{F}_q^{\star} = \mathbb{F}_q \setminus \{0\}$. The number of such solutions of $Ax = 0$ then is equal to $(q-1)^{t-1}$.

6.1. **Structure of matrices corresponding to cycles of associate Tanner graphs.** .

For bipartite graphs with cycles, there is an interesting structure in the corresponding parity check matrix. The underlying structure of the matrix corresponding to cycle in Tanner graph can be summarized in the following way [12].

**Definition 6.** A $2g$ cycle matrix $M$ is a $g \times g$ matrix over $\mathbb{F}_q$ satisfying the following conditions:

(1) There are exactly two non zero elements ($\in \mathbb{F}_q$) in each row of $M$
(2) There are exactly two non zero elements ($\in \mathbb{F}_q$) in each column of $M$
(3) For any square sub-matrix $N \subset M$, $N$ does not satisfy the previous two conditions simultaneously.

**Example 7.** A $4-$ cycle, along with the associated matrix representation is shown in Figure.4. The code is defined in $\mathbb{F}_4$.

**Example 8.** A $6-$ cycle matrix representation defined over $\mathbb{F}_4$ is given by $M_3$.

**Example 9.** A bipartite graph and cycle representation of a $8-$ cycle matrix is shown in Figure.5. The code is defined over $\mathbb{F}_4$.

$$
\begin{aligned}
M_2 &= \begin{pmatrix} 1 & 1+\alpha \\ \alpha & 1 \end{pmatrix} \\
M_3 &= \begin{pmatrix} 1+\alpha & 1 & 0 \\ 0 & \alpha & 1+\alpha \\ 1+\alpha & 0 & \alpha \end{pmatrix}
\end{aligned}
$$

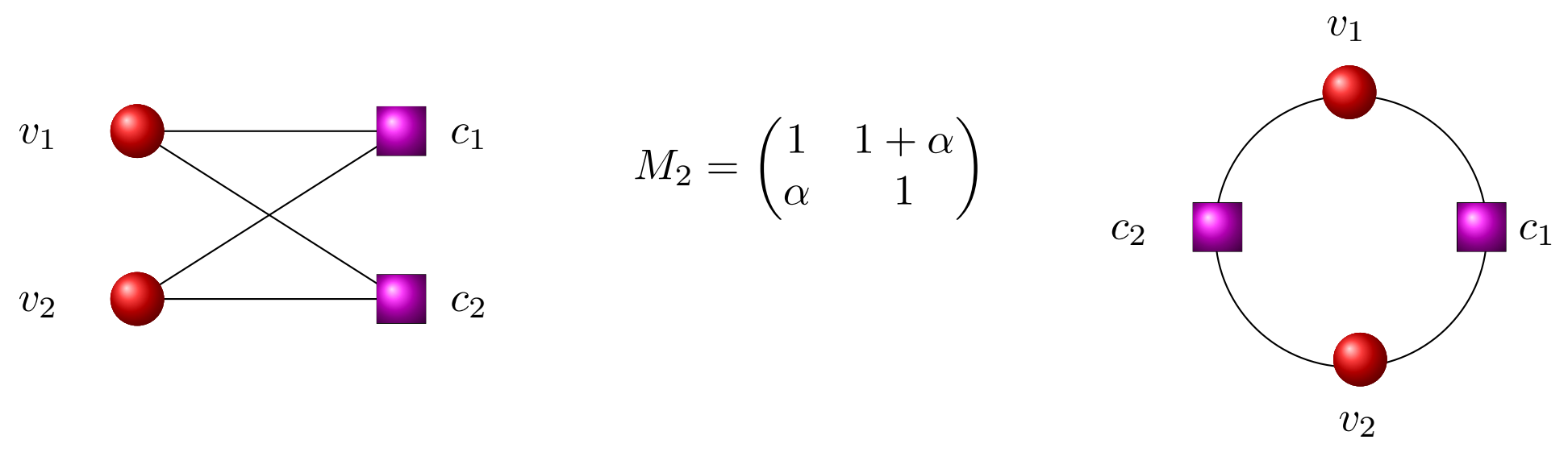

**Figure 4**. Single cycle: The smallest cycle ($4-$ cycle) and its matrix representation are shown. Each row and column has exactly 2 non zero elements. The matrix is defined over $F_4$. The primitive polynomial generating $\mathbb{F}_4$ is $p(x) = 1 + x + x^2$. The primitive root is $\alpha$ and the elements of $\mathbb{F}_4$ are $\{0, 1, \alpha, 1+\alpha\}$.

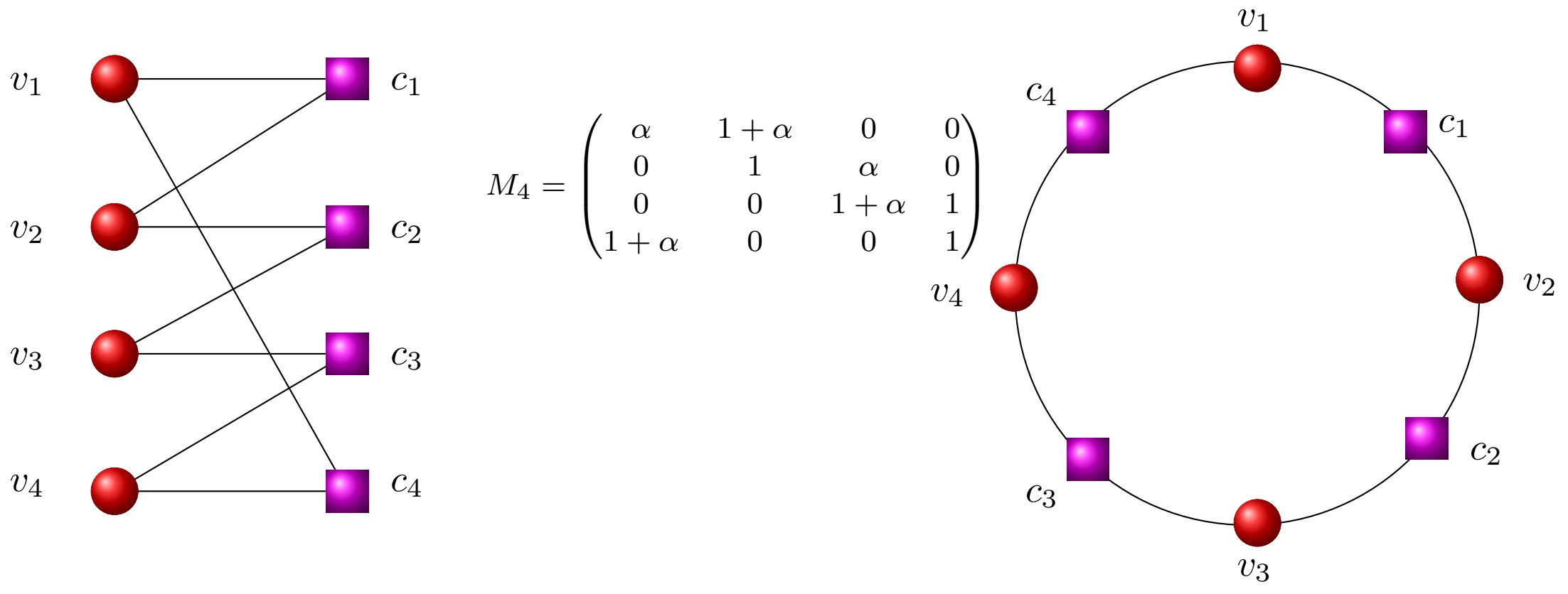

**Figure 5**. A $8-$cycle graph and the corresponding matrix representation are shown. All the columns and rows of this square matrix have exactly 2 non zero elements. The matrix is defined over $F_4$. The primitive polynomial generating $\mathbb{F}_4$ is $p(x) = 1 + x + x^2$. The primitive root is $\alpha$ and the elements of $\mathbb{F}_4$ are $\{0, 1, \alpha, 1+\alpha\}$.

It may be noted that, for sub-graphs which are cycles, the corresponding sub-matrix of the parity check matrix is square. We will soon see (in the following sections) that, for other structures such as joint cycles, the associated sub-matrices no longer stay square.

In the case of single cycles, Poulliat (2006) et al have recently established the following result [9].

**Theorem 10.** *If $\alpha_0, \alpha_2, \ldots, \alpha_{T-1}$ are independent random variables which take the values $1, 2, \ldots, q-1$ independently and with equal probabilities, then $P_C$, the probability that there exist solutions other than the non trivial (all zero) is given by,*

$$P_C = \frac{1}{q-1}. \tag{3}$$

6.2. **Matrix structure of associated graphs with joint cycles.** When the sub-matrix corresponding to the sub-graph (of a Tanner graph) has the following structure, that correspond to union of two joint cycles.

(1) Every column has exactly two non zero elements. The non zero elements are the non zero elements of the field $\mathbb{F}_q$ over which the code is defined.
(2) Exactly two rows have 3 non zero elements whereas the remaining rows have exactly 2 non zero elements. The non zero elements are the non zero elements of the field $\mathbb{F}_q$ over which the code is defined.

**Example 11.** A matrix representation of a configuration with union of two joint cycles is shown in Figure.6 along with the bipartite graph and cycle representations.

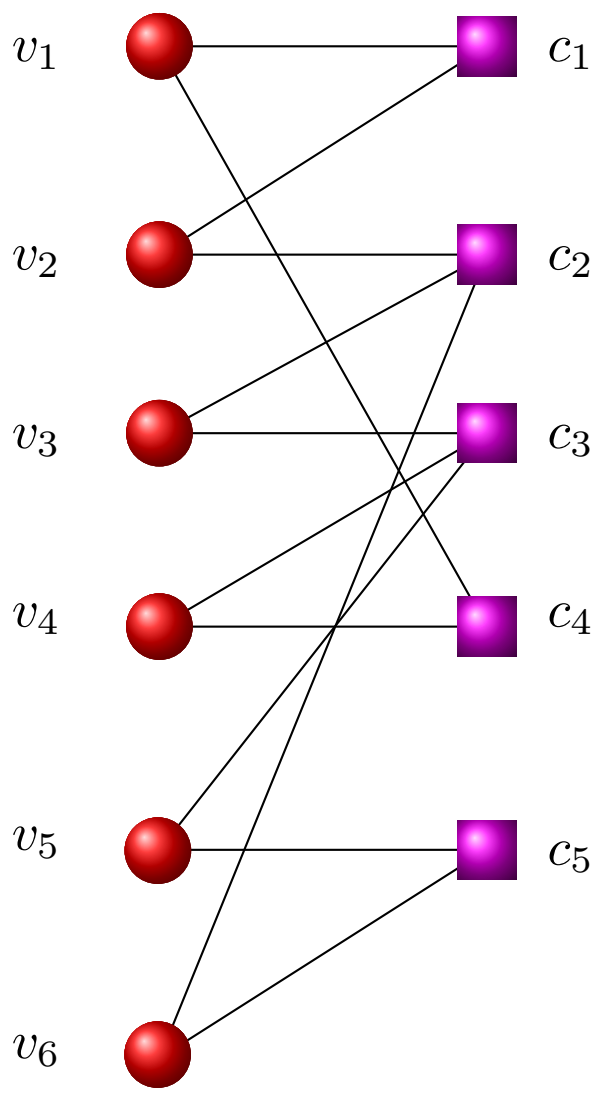

$$\begin{pmatrix} \alpha & 1 & 0 & 0 & 0 & 0 \\ 0 & 1 & \alpha & 0 & 0 & 1 \\ 0 & 0 & 1+\alpha & \alpha & 1 & 0 \\ 1+\alpha & 0 & 0 & 1+\alpha & 0 & 0 \\ 0 & 0 & 0 & 0 & \alpha & 1+\alpha \end{pmatrix}$$

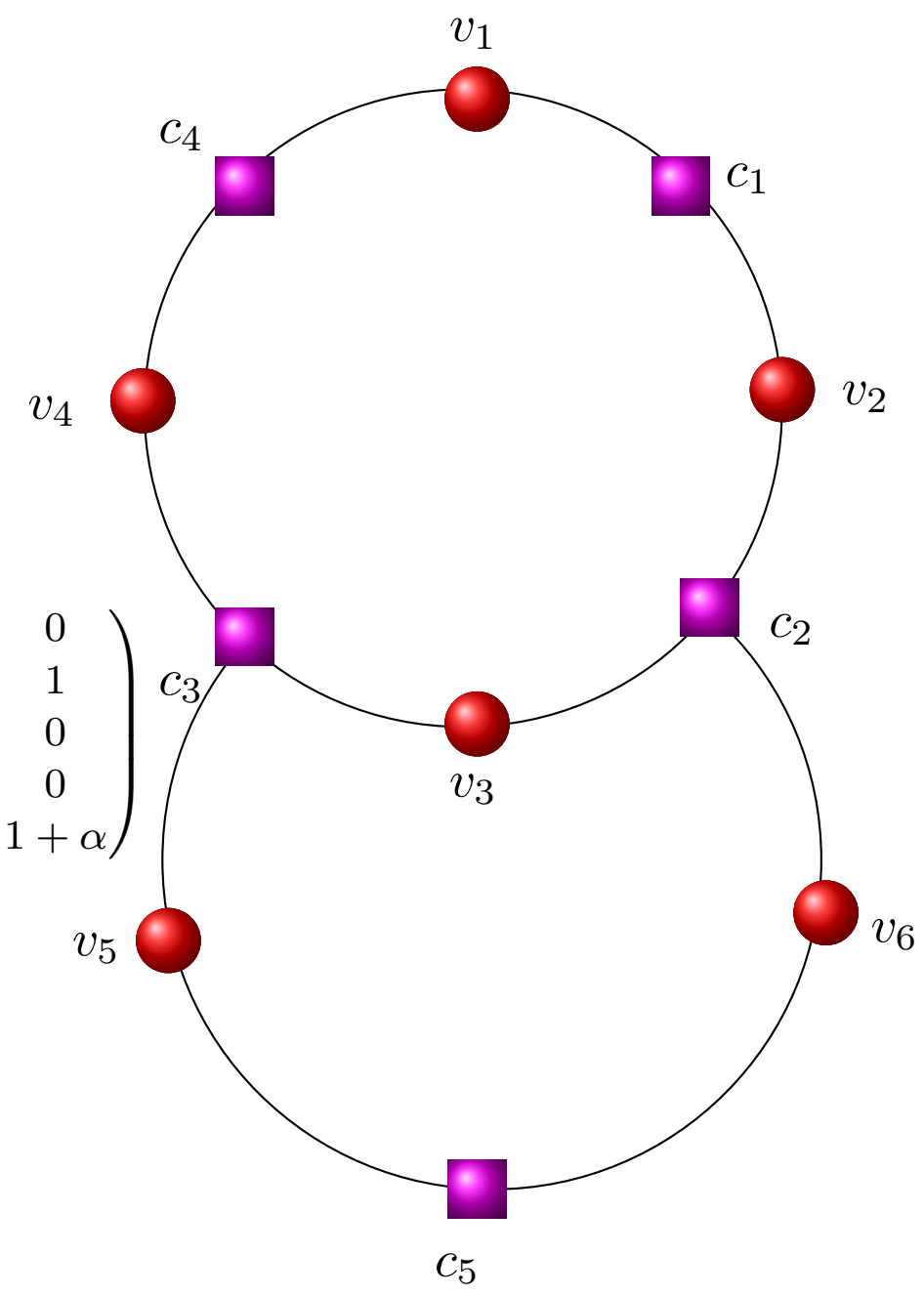

**Figure 6**. Joint cycles: All the columns of the matrix has 2 non zero elements. Two rows have 3 non zero elements, while the remaining rows all have 2 non zero elements. The matrix corresponding to this system is rectangular. The matrix is defined over $F_4$. The primitive polynomial generating $\mathbb{F}_4$ is $p(x) = 1 + x + x^2$. The primitive root is $\alpha$ and the elements of $\mathbb{F}_4$ are $\{0, 1, \alpha, 1+\alpha\}$.

In the case of joint cycles, the matrix assumes a rectangular shape (as against a square matrix representation of single cycle). The number of columns for such matrices are higher than the number of rows. In other words, the number of equations are less than the number of variables.

Matrix representation correspond to union of more than one cycles is just an extension of the 2 cycle union. The matrix stay rectangular as it was the case for union of two cycles.

6.3. **Union of joint cycles.** As discussed earlier, the union of joint cycles correspond to a linear system of equations, with certain structure. The linear system $Ax = 0$ with $A$ rectangular. Since the number of variables are more than the number of equations, there would exist multiple solutions.

This is a linear system of equations of the form $Ax = 0$, where $x$ is the solution to the kernel (null space) of $A$. The number of nonzero solutions of $x$ satisfying $Ax = 0$ is equal to $(q-1)^{T-\text{rank}(A)}$. In the case of two joint cycles in a graph, the $rank(A) = T - 1$, which lead to a total of $(q-1)$ solutions. In general, for $t$ joint cycles, $rank(A) = T - t + 1$ and consequently, a total of $(q-1)^{T-(T-t+1)} = (q-1)^{t-1}$ non trivial (non zero) solutions.

A special case of joint cycles is *chains of cycles*.

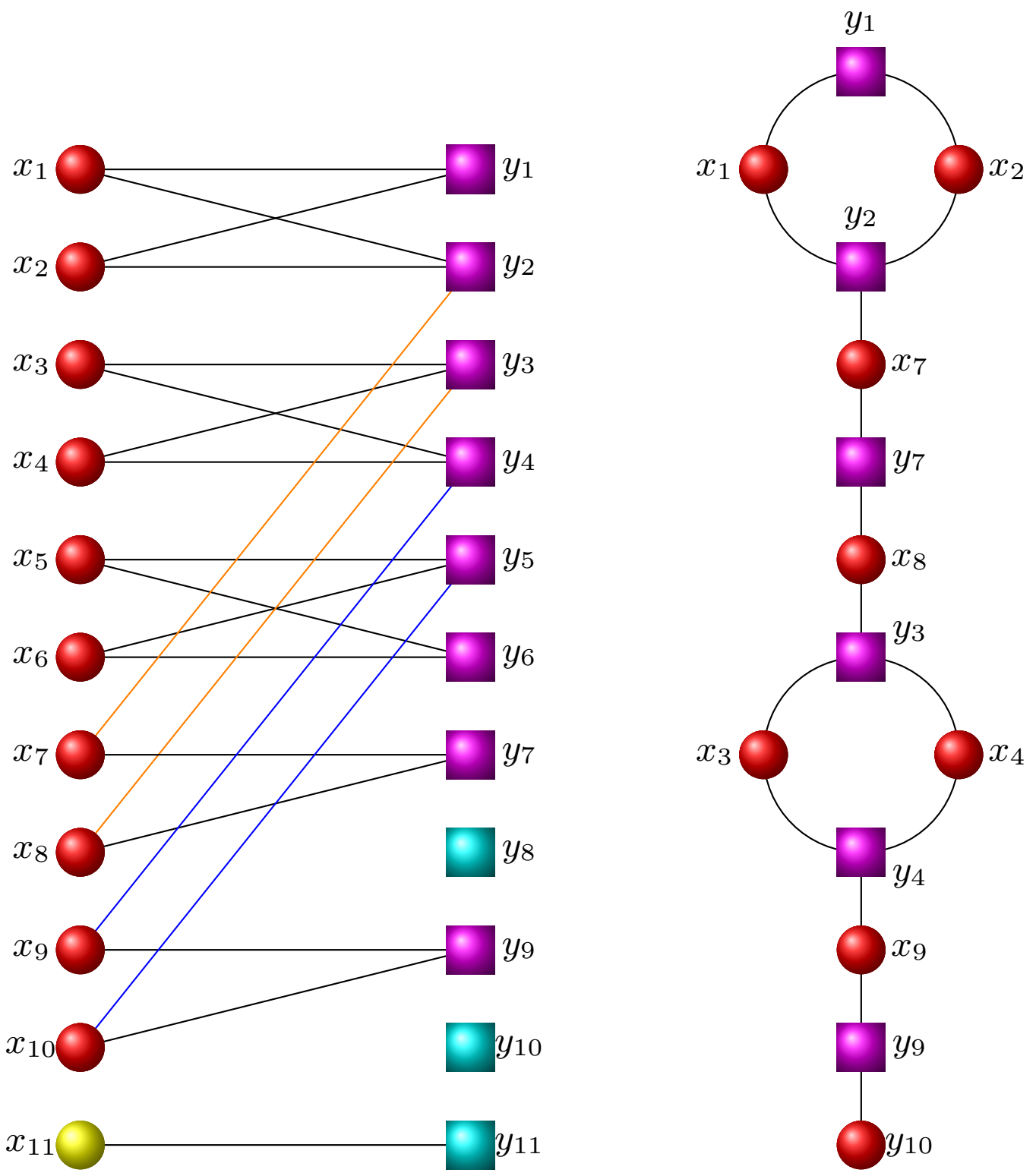

**Figure 7**. Chain of cycles: On the left is the bipartite graph and on the right shown is the cycle view of the graph. Both represent the same graph.

6.4. **Chain of cycles.** In the case of chain of cycles, two or more cycles are connected by a link involving series of check nodes and variable nodes. The connecting link will always have one variable node more than the number of check nodes. An example of chain of cycles is shown in Figure.7.

Let us consider bipartite graph with variable node degree equal to 2. We partition the graph into two sub-graphs $S_1$ and $S_2$. The set $S_1$ contains disjoint cycles, whereas $S_2$ consists of chains. Let $n_1$ be the number of variable nodes in $S_1$ and $m_2$ is the number of check nodes. In $S_2$ we have chains. Several combinations are possible here (They are yet to be listed here). Let $n_2$ be the number of variable nodes in $S_2$ and let $m_2$ denote the number of check nodes in this partition.

Clearly $n_1 = m_1$. There are different configurations possible for chains. The worst case scenario would be $m_2 = \lceil \frac{n_2}{2} \rceil$. The best case (least number of check nodes in $S_2$ would be 1. In the latter configuration, all the variable nodes forming the chain are connected to check nodes from $S_1$.

Let $a$ be the average number of degree 2 variable nodes in a chain. Let $C$ denote the number of chains. We are interested to find a relationship between $n_2, C$ and $a$. Let $a_i$ be the number of degree-2 variable nodes in chain $i$. The number of check nodes in chain $i$ is then equal to $a_i - 1$. The total number of variable nodes in chain is $n_2$. Similarly,the total number of check nodes in chain is $n_2$. That is $n_2 = \sum_{i=1}^{|C|} a_i$ and $m_2 = \sum_{i=1}^{|C|} a_i - 1$

$$
\begin{aligned}
a &= \frac{\text{Total number of variable nodes counted in all chains in graph}}{\text{Total number of chains in the graph}} \\
&= \frac{n_2}{|C|} \\
|C| &= \frac{n_2}{a} = \frac{m_2}{a-1} \\
a(m_2) &= (n_2)(a-1) \\
n_2 &= m_2 \frac{a}{a-1}
\end{aligned}
$$

When $C = 1$, it is a special case (unique cycle).

## 7. Simulation

In order to verify the existence of the (valid codeword) configurations discussed thus far, we have performed some simulations using non binary LDPC codes. The simulation setup consisted of the following. LDPC codes defined over $\mathbb{F}_4$ is used. The decoding algorithm chosen is belief propagation (BP), under Additive white Gaussian noise (AWGN) channel. Both $(2,3)-LDPC$ and $(2,4)-LDPC$ schemes are considered. We performed simulations for three different codelengths ($n = 900, 1800$ and $9000$). Channel parameters are chosen such that, we are in error floor region (of the bit error performance). We considered error events less than $50$. We have considered structure of codeword configurations corresponding to error events. Random permutations and recursive strategy to avoid loops with multiple edges are used. With limited number of simulations, we have discovered some codewords with cycle configurations. We have also identified certain codeword configurations with few other interesting structures. So far, we have not identified any codeword configurations with union of joint cycles, but we expect the need for longer simulation runs to identify more interesting structures. In [9], they have identified configurations with union of 3 joint cycles. However they have not reported any cases of union of 2-joint cycles. We are performing more extensive simulations to identify some of these configurations.

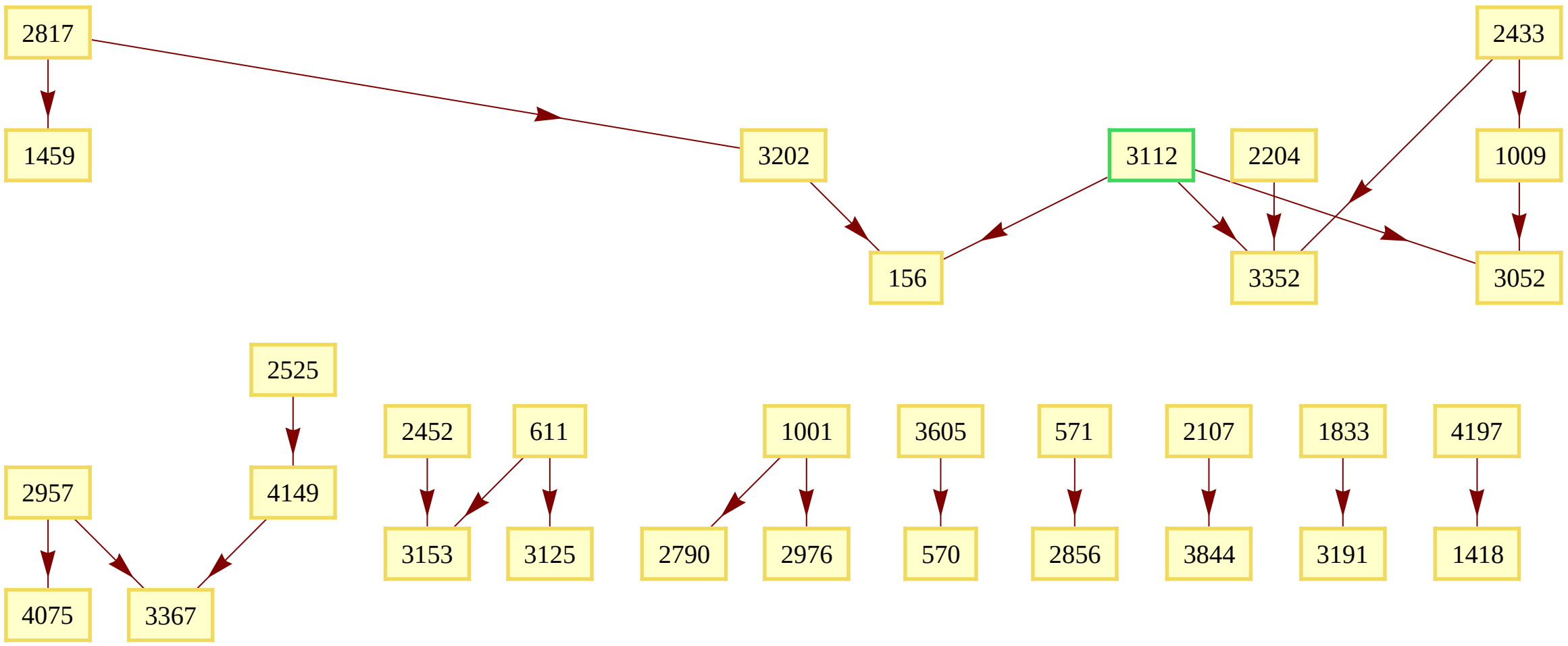

**Figure 8**. Simulation result: Configuration with one cycle:

## 8. Discussion on bounds

In the case of binary LDPC codes, the minimum distance of a codeword length correspond to the smallest cycle in the graph representation of the code. The length of the smallest cycle in a graph is referred to as the girth[13]. Since a cycle of length $t$ give rise to a codeword of weight $t$ (just set the variable nodes involved in the cycle to $1$ and all remaining variables to $0$ [8]). the girth of Tanner graph then directly infer the minimum distance. If the length of such cycle is logarithmic, then that lead to logarithmic minimum distance. In other words, logarithmic bounds on girth then imply logarithmic bound on minimum distance.

For non-binary codes, the result is not immediate. For LDPC codes over $\mathbb{F}_q, q > 2$, cycles correspond to codewords with some probability. This probability decreases with increase in alphabet size ($q$). However, because of the higher degrees of freedom available in a non binary code, such codewords can almost always be avoided by appropriate choice of edge labels (In binary case, there is no degree of freedom available since $1$ is the only non-zero value). Even in the case of $2-$regular codes, cycles which correspond to valid codewords, can be avoided in non-binary case. In binary codes, a codeword with a cycle configuration cannot be avoided.

On the other hand, codeword configurations which are joint cycles cannot be avoided. This in a way serve as the deciding parameter on the minimum distance of non binary codes. We can obtain a bound on logarithmic $d_{\min}$ using similar approach as that of binary codes.

8.1. **Girth bound adapted to nonbinary codes.** For binary codes, the bound on logarithmic $d_{\min}$ can be be proved as follows. We know that, a cycle of length $t$ give rise to a codeword of weight $t$ (simple assignment of $1$ to all the variable nodes in the cycle and $0$ to every other variable nodes). Consider a tree structure of the graph (a tree stemmed from a check node), where we assume that, there are no cycles up to (maximum) depth $t_d$. Then, there is a cycle at depth $t_d + 1$. In other words, the maximum depth of the tree graph is $t_d$. The tree depth $t_d$ is logarithmic in the codelength $n$ if the number of degree-2 variable nodes ($n_2$) is more than the number of check nodes ($m$) [14] [15].

We can adapt the bound strategy to nonbinary codes, with a simple exception in the graph. Here, we begin with a cycle. We construct a tree-like graph stemmed from the check node of the cycle (Remember that the graph constructed thus far is not a tree-graph in the strict sense. What we constructed is essentially a tree structure which stemmed down from a check node of a cycle). Recall that, a codeword with a corresponding cycle configuration can be avoided in the non-binary case. Thus, a single cycle in the graph considered is not corresponding to a codeword. Let $t_d$ be the depth of the graph, until when we have no further cycles. Then, at depth $t_d + 1$, we have another cycle and this new cycle form a joint cycle with the earlier cycle (from which the graph stemmed). The union of two joint cycle indeed form a valid codeword and this configuration cannot be avoided. By same argument as the girth bound for binary codes, we can arrive at the bound on logarithmic $d_{\min}$ as $t_d = O(\log n)$ if $n_2 > m + 1$. Note that the only difference between this bound and that of the binary code (girth bound) is the additional term 1, which is due to the fact that, there is already a cycle in the graph within graph of depth $t_d$. Asymptotically (as $n \to \infty$), the two bounds (binary and nonbinary) approach the same bound.

## 9. Concluding remarks and scope for further work

In this report, we have discussed the results of our preliminary study on the necessary conditions for the linear minimum distance growth of non-binary LDPC code ensembles. We have studied some specific configurations which lead to low weight codewords. Our simulations have helped to identify some of these configurations. It is hoped that, with extensive simulations, more such configurations could be identified. While girth of the Tanner graph directly provide a measure of minimum distance for binary codes, adapting the girth bound to joint cycle configuration help us to obtain a bound for logarithmic minimum distance, in the case of nonbinary codes. Asymptotically, these two bounds indeed behave similar. Because of the increased number of freedom available in nonbinary codes (to choose edge labelling), it is perhaps possible to achieve improvements on these bounds with appropriate edge labelling.

## References

[1] R. Gallager, Low Density Parity Check Codes, MIT Press 1963.
[2] T.Richardson, R.Urbanke, Modern Coding theory, Cambridge University Press, 2007.
[3] J.H. Van lint, Introduction to Coding Theory,Third Edition, Springer 1998.
[4] T.Moon, Error correction coding: Mathematical methods and Applications, Wiley International 2005.
[5] V.Guruswami, "Iterative Decoding of Low Density Parity Check Codes (An Introductory Survey)",Electronic Colloquium on Computational Complexity, Report No.123 (2006)
[6] R.Tanner, "A recursive approach to low density Codes",IEEE Trans. on Information Theory,Vol 27, Sept 1981,PP:533-547
[7] Iryna Andriyanova, "Analysis and Design of a Certain Family of Sparse-Graph Codes: TLDPC Codes",PhD thesis INST Paris 2006",
[8] C.Di, "Asymptotic and Finite length analysis of Low density Parity Check Codes", PhD thesis Écolé Polytechnique Fédérale de Lausanne (EPFL), 2004
[9] C.Poulliat, M.Fossorier, D.Declercq, "Design of non binary LDPC codes using their binary image:algebraic properties", ISIT 2006, Seattle, USA, July 19-14, 2006, PP:93-97
[10] J.H.Hodges, "The Matrix Equation AX=B in a Finite Field", American Mathematical Monthly, Vol.63, No.4, April1956, pp. 243-244 (Accessible from http://www.jstor.org/stable/2310349)
[11] V.F.Kolchin, Random Graphs, Encyclopedia of Mathematics and it Applications: Vol.53, Cambridge University Press, 1999
[12] R.Chen,H.Huang and G.Xiao, "Relation Between Parity Check Matrices and Cycles of Associated Tanner Graphs", IEEE Comm.Letters, Vol. 11, No.8, August 2007.
[13] R.Diestel, Graph Theory, 2nd Edition, Springer International Edition, 2000
[14] X.Hu, "Low-delay low-complexity error-correcting codes on sparse graphs", PhD thesis, Écolé Polytechnique Fédérale de Lausanne (EPFL), 2002
[15] N.Alon, S.Hoory and N.Linial, "The Moore bound for irregular graphs", *Graphs Combin.*, 18:53-57,2002

*E-mail address*: `rethnakaran.pulikkoonattu@epfl.ch`